# Dynamic Facial Expression of Emotion Made Easy

Joost Broekens[1], Chao Qu, Willem-Paul Brinkman

Man-Machine-Interaction department, Delft University, The Netherlands.
Joost.broekens@gmail.com, c.qu@tudelft.nl, w.p.brinkman@tudelft.nl

**Abstract.** Facial emotion expression for virtual characters is used in a wide variety of areas. Often, the primary reason to use emotion expression is not to study emotion expression generation per se, but to use emotion expression in an application or research project. What is then needed is an easy to use and flexible, but also validated mechanism to do so. In this report we present such a mechanism. It enables developers to build virtual characters with dynamic affective facial expressions. The mechanism is based on Facial Action Coding. It is easy to implement, and code is available for download. To show the validity of the expressions generated with the mechanism we tested the recognition accuracy for 6 basic emotions (joy, anger, sadness, surprise, disgust, fear) and 4 blend emotions (enthusiastic, furious, frustrated, and evil). Additionally we investigated the effect of VC distance (z-coordinate), the effect of the VC's face morphology (male vs. female), the effect of a lateral versus a frontal presentation of the expression, and the effect of intensity of the expression. Participants (*n*=19, Western and Asian subjects) rated the intensity of each expression for each condition (within subject setup) in a non forced choice manner. All of the basic emotions were uniquely perceived as such. Further, the blends and confusion details of basic emotions are compatible with findings in psychology.

## 1 Introduction

In this article we propose an easy-to-use method to generate dynamic affective facial expressions for virtual characters. The mechanism is easy to implement and downloadable code examples are provided. We evaluate the recognition accuracy of the generated expressions as well as how this accuracy varies under the influence of four experimental variables. These variables include: what is the effect of *distance* on the perception accuracy; what is the effect of variation in the virtual character's *face morphology*, what is the effect of a *lateral versus frontal* presentation of the character's face, and what is the effect of expression *intensity*? Understanding the effect of these variables is relevant in particular for affective virtual characters in 3d environments where the user or character can move around.

---

[1] Corresponding author.

Emotions and affect play an important role in human-computer interaction (Hudlicka, 2003; Picard, 1997; Picard & Klein, 2002). A common role found in many application areas as well as scientific studies is when emotion and affect are modeled in a virtual character or agent (Gratch et al., 2002). In this case, the virtual character is modeled as a "virtual human" including affective abilities such as user emotion recognition, artificial emotion, personality or mood, and emotion expression towards the user. Sometimes all of these abilities are modeled in one character, sometimes only one or several.

The main reason to include models of emotion in these characters is because some form of social interaction is needed between the character and the human, and emotion facilitates social interaction (Vinayagamoorthy, Steed, & Slater, 2005). Reasons to do so include applied ones such as enhancing believability and/or realism for the character or enhancing entertainment value (Vinayagamoorthy, et al., 2005), as well as more theoretical ones such as the study of emotion by using computer models (see e.g., (Broekens, 2010; Gratch, Marsella, & Petta, 2009)).

Believability and realism is important for example in the domain of virtual reality training when a user is trained on a real-world task and interaction with virtual characters is part of the training (Core et al., 2006; Gratch & Marsella, 2001; Traum, Marsella, Gratch, Lee, & Hartholt, 2008). Related to this is the area of tutor agents (aka pedagogical agents) in which virtual characters are used as a conversational interface component helping a user to perform a learning task. Key here is that affective abilities help these characters to communicate in a for humans natural way (Graesser, Chipman, Haynes, & Olney, 2005; Heylen, Nijholt, Akker, & Vissers, 2003; Heylen, Vissers, op den Akker, & Nijholt, 2004), for example using models of empathy and the expression thereof (Hall, Woods, Aylett, Newall, & Paiva, 2005; McQuiggan & Lester, 2007; Picard & Klein, 2002; Prendinger & Ishizuka, 2005). In VR training and tutor environments a related goal is to enhance the level of presence (the sense of being there) (IJsselsteijn, Ridder, Freeman, & Avons, 2000) by using socially expressive characters (Vinayagamoorthy, et al., 2005). The idea is that having a higher level of presence ultimately helps carrying over the training effect to a real situation, or enhances the level of entertainment.

Entertainment value can be increased using affective abilities, for example by making a virtual character in a game (aka NPC, non-player character) more flexible with regards to their behavior as well as react in more human-like ways (Hudlicka, 2008; Hudlicka & Broekens, 2009). Further, games can be made more interesting and fun by making them affect-adaptive, i.e., sense and model user affect (or affect-related factors such as frustration, excitement, challenge and curiosity) to adapt the game in order to enhance the level of engagement and fun as perceived by the user (Gilleade & Dix, 2004; Yannakakis & Hallam, 2006).

Finally, modeling affective abilities can make us understand more about emotions per se, for example by modeling how emotions arise from cognitive processes (Gratch, et al., 2009), or how different sources within a virtual character (e.g., mood and emotions) influence its affective facial expressions and how blended emotions should be modeled (Ochs, Niewiadomski, Pelachaud, & Sadek, 2005; Rosis, Pelachaud, Poggi, Carofiglio, & Carolis, 2003).

All of the above mentioned affective abilities are equally important for the generation of believable, human-like virtual characters (and robots). However, when user interaction is required a valid and reliable way of expressing affect is needed. The expression should be recognized by the human as intended (joy must be perceived as joy, not disappointed or evil). Whenever the expression of social signals is needed in a study or application the need for validated emotional expressions exists, and this validation involves major effort.

Validity and reliability of facial affective expressions have been addressed in the recent past (Bartneck, 2001; Bartneck & Reichenbach, 2005; Carolis, Pelachaud, Poggi, & Steedman, 2004), resulting for example in an MPEG4 standard for influencing faces according to the Facial Action Coding System (FACS) model (Ekman & Friesen, 2003). In the related work section we will review the current state of the art in more detail. For now it suffices to state there are three important issues to take into account when assessing recognition accuracy of affective expressions:

1. *Include blend emotions in the evaluation.* It is important to not only test the set of basic emotions, but also blends, because this gives important information about the generative power of the model. This is relevant in particular when the emotional state of the virtual characters is dynamic (e.g., represented by factors such as pleasure, arousal and dominance) (Breazeal & Scassellati, 2000; Broekens & DeGroot, 2004).
2. *Include confounding variables* such as distance to the virtual character, intensity levels of the expression, viewing angle towards the virtual character, and reliability of the method. As many of the virtual characters are evaluated in a typical frontal view, often with pictures of the extreme, at the same distance and without manipulating the intensity, it is difficult to generalize the validity of the expression model. This becomes a bigger problem if these virtual characters must be used in actual 3d worlds (VR training, games, etc.). Now the user and virtual character are mobile, and there is no guarantee that a frontally perceived expression of joy is also recognized as joy in a lateral view.
3. *Use a non-forced choice intensity-based rating scheme.* In many studies, the validity (recognition correctness) is assessed either through a forced choice scheme, or in a comparison scheme. In the first case, the user has to make a choice, meaning that a polarized view will arise about the recognition rate especially in the absence of blended emotional expressions as test stimuli. A little bit of joy will quickly be rated as joy if the alternatives are surprise, anger, sad, disgust and fear. People are inherently good at making forced choices, especially when they know what is expected, and this will blow up the recognition accuracy. In the comparison scheme, subjects are asked to compare virtual faces with human faces. This is a good way of figuring out if what is modeled looks like the human, but problematic when interested in affective attribution to virtual affective faces and the details between these attributions, especially when many different facial expressions are possible.

In this article we present a mechanism for the generation of affective facial expressions in virtual characters that is easy to use and implement. Our method does not need specific "affective" tools or programs. It can be repeated in any combination of 3d modeling environment and VR development tool. To put our contribution in perspective: we do not claim a breakthrough in realism, nor do we claim a completely novel way of generating affective facial expressions. There are more believable expressions available, for example in animation. There are more advanced mechanisms available, such as those that express sophisticated social signals through the face. Our contribution is a method that is easy to use, that enables the dynamic generation of affective facial expressions in virtual characters, and that is evaluated according to a very high standard of evaluation. For many researchers who use virtual characters the primary aim is not the study of affect expression. We feel that the lack of a thoroughly validated, easy-to-use method is limiting progress. It means that each researcher needs to develop affect expression in virtual character from scratch based on data from papers, and has the burden to evaluate the expressions.

Our method is based on a muscle based FACS system, where the muscle attachments are modeled in the 3D head as anchors with a region of interest of vertices, and the FACS influences are modeled as a vector pulling the regions in a particular direction. Separating these two things enables the quick re-instrumentation of a new virtual character, because the vectors and expression intensities can be re-used.

To show the validity of the expressions generated with our method we tested the recognition accuracy for 6 basic emotions (joy, anger, sadness, surprise, disgust, fear) and 4 blend emotions (enthusiastic, furious, frustrated, and evil). Additionally we investigated how the perception of expressions changes under the influence of 4 experimental variables: virtual character face morphology (male vs. female), distance (1 vs. 3 meters), geometric intensity (high vs. low), and viewing angle (frontal vs. lateral). Expressions were generated and presented in the 3d world (no pictures or movies), with realistic onset, hold and decay timings. Participants ($n$=19) – a mix of Western (n=13, mostly Dutch) and Eastern (n=6, mostly Chinese) subjects – rated the intensity of each expression for each condition (within subject setup) in a non forced choice manner. Our results show that the proposed mechanism is easy to use and produces valid expression that are in line with psychological findings.

In the next section we discuss the state of the art of affective facial expression in more detail. We then describe the method we have used including the actual FACS encodings, weights, etc. Then we introduce the experimental setup to test the four main research questions and the validity of the resulting affective facial expressions. Finally we discuss our experimental results.

### 1.1 Related work

Implementing behavioral social cues in virtual characters is a subtle process and positive results can be highly dependent on the context of the virtual character, the other behaviors of the character, the user, as well as the task the user is doing (for reviews see (Vinayagamoorthy et al., 2006; Vinayagamoorthy, et al., 2005)). As such one can debate if singling out one particular channel of communication, in our case affective facial expressions, is a good way to take in the first place. For example, (Rosis, et al., 2003) present a framework for blending emotional facial expressions that can also be used to integrate the different communicative functions of the face. (Pelachaud & Poggi, 2002) also proposed a related framework for the modeling and blending of the different communicative functions of the face. In an attempt to better structure such expressions and link them to different affective sources (such as an emotional state, display rules) (Carolis, et al., 2004) introduce the Affective Presentation Markup Language (APML). Finally, Lance and Marsella (Lance & Marsella, 2008) show how emotionally expressive gaze can be embedded into a VR character model and mixed with emotions to support the expression thereof.

Having said that, facial expression of emotion is a powerful modality to express emotion (Elfenbein & Ambady, 2002). It makes sense to try to understand how to model these expressions, so that humans are able to recognize these in virtual characters. In this section we review studies that have tried to do so explicitly. Although many of the studies reviewed not only present a model for expression but also a model for the representation of the affective state or personality factors (e.g., (Arellano, Varona, & Perales, 2008), we focus in our review on the experimental evaluation of the facial expressions in these studies.

### 1.1.1 Intensity effects

In a study on the perceived distinctiveness, intensity and convincingness (naturalness) of images of basic facial expressions, intensity correlated with convincingness (Bartneck, 2001). This shows the importance to study different intensity levels. In this study, no blended emotions were used, and the rating was done with a forced choice scheme. It is interesting to see if these results are similar if the rating scheme is a non-forced choice one, as in our study. We address this in the results section.

In a later experiment the influence of geometrical intensity on perceived intensity was tested (Bartneck & Reichenbach, 2005). This is the same setup as we use (see later) for intensity manipulation. It was shown that geometrical intensity can successfully modulate perceived intensity of static faces (non-3d virtual character). They used a Likert scale non forced choice method. However, there was no use of blended emotions or disgust. Problems with the perception of fear were reported, while happiness was well recognized. These results show that geometrical intensity can be used to manipulate perceived intensity. As this is a simple and repeatable mechanism, we use it to implement intensity.

### 1.1.2 Blended emotion recognition

In a fairly large user study (n=75) (Arellano, et al., 2008) investigate the perception of 16 basic and blended expressions generated based on a system comparable to ours (i.e., FACS-MPEG4 based, linked to a dynamic affective state). Users rated using a forced choice method in which they had to choose out of 8 different "basic" emotions from the PAD affect space extremes (combinations of high and low Pleasure, Arousal, and Dominance). The stimuli consisted of static images and videos of the frontal view of the extremes of the chosen expressions, and there was only one virtual character morphology tested. The expression timing used in the videos is mentioned but not related to psychological findings. The study gives several interesting insights with respect to blended and dynamic emotions. First, blended emotions are difficult to recognize as such, and people tend to rate them as belonging to an extreme. This might be a result of the forced choice method, though. If a subject is forced to choose, then the tradeoff subjects make is difficult to measure and as such the blend will be less visible in the ratings. Second, basic emotions, except fear and disgust, are well recognized in this forced choice scheme. Third, eyes are important and distracting if they don't move at all, adding to the evidence that eye movement should be taken into account for emotional facial expressions (Lance & Marsella, 2008).

(Bevacqua, Mancini, Niewiadomski, & Pelachaud, 2007) present a sophisticated method to model basic and complex emotions, by which they mean, e.g., the ability to model and express masked and superposed affective expressions. This framework also allows to incorporate multiple sources for the expression as well as other means of expression (based on (Ochs, et al., 2005)), and is a richer model for blending than ours (which is simply based on a linear combination of muscle movement, see later sections). Our model allows for superposition, not masking, as we use a muscle based linear addition (see later sections). Related to this work, (Niewiadomski & Pelachaud, 2007) show how to create blended emotional expressions using a fuzzy rule-based system. This is an alternative approach to ours, where the blended emotions are simply linear combinations of the facial expressions. In another study, the same group has analyzed to what extent these complex generated emotions match the emotions in the original video clip (EmoTv) by asking subjects (Buisine et al., 2006) how similar the real video clip was to the expression generated by their system in a non-forced choice manner. The results were promising, although the basic generated expression was seen most similar to the blended expressions in the video, meaning that blended emotions were often perceived as basic ones. In addition to that, the addition of an audio channel (original audio from video clip) to the stimulus influenced this similarity measure, showing that the recognition of blended emotions can be strongly dependent on context variables .

Interestingly, (Noël, Dumoulin, & Lindgaard, 2009) report that textual context did *not* influence the recognition of facial expressions of extreme (static) faces (human and virtual) in a forced choice paradigm. However, as the second authors used a forced choice method, while the first authors do not, and as the second tested basic emotions and not blended, these studies cannot be compared to each other with respect to their conclusion about the influence of content on the perception of facial expressions. Further, they had high variation in affective attributions to the neutral face (consistent with psychological literature).

**1.1.3 Static versus dynamic expressions**
(Kätsyri & Sams, 2008) did not find a difference between static versus dynamic artificial expressions, in an experiment that was otherwise similar to ours (no forced choice and including a naturalness rating). However, their dynamic stimulus ends with the extreme of the expression, not with the offset like ours. Further, the stimulus is continuously presented until the rating is finished, and the stimulus is not presented in a VR context but isolated. Further, they do not address blended emotions, and their expressions are implemented manually and not based on a dynamic generation model for expressions, like ours. Therefore, although their results seem promising in the sense that static expressions can be equated with dynamic expressions, we feel additional study is warranted to really draw such a conclusion in a more realistic setting (i.e., complete expressions, timings based on literature, blended emotions).

(Noel, Dumoulin, Whalen, & Stewart, 2006 ) show that there is little impact on recognition of static versus dynamic expressions. They use a forced choice rating scheme, and do without blended emotions. As mentioned earlier, the use of a forced choice rating scheme can impact the variability of the ratings, and it is therefore difficult to conclude that there is no difference between static and dynamic expressions based on forced choice ratings.

### 1.1.4 Morphologies and context

All of the reviewed studies remove the to-be-evaluated faces (stimuli) from the natural context of a 3d world. The effect of distance or angle of view on the perception of the expression has hardly been studied. Further, most of the studies use close-up static images, or video segments, not the actual 3d model in a 3d world. Finally, when a method is evaluated in these studies, it is common to validate the method using only one morphology for the virtual character's face. We feel that it is necessary to validate using at least two morphologies for the simple reason that a method should be applicable to (and therefore validated on) multiple facial morphologies. In our study we explicitly introduce the variables distance, viewing angle, and face morphology to investigate the effects of these variables on the user perception of the expressions.

### 1.1.5 Rating schemes

As is clear from our short review of the state of the art, many studies use forced choice rating schemes, without explicitly asking for the naturalness of the expression. We feel forced choice rating has a couple drawbacks in light of evaluating validity and reliability of expressions. First, forced choice rating forces people to make polarized choices resulting in better matches. Second, it shadows potentially subtle differences (e.g., when studying perceptual differences between static and dynamic faces). Third, it introduces a lot of variance when rating blended emotions (people have to choose between the components that make the blend), or it forces the blended emotion in a dominant basic category (if one of the components of the blend is slightly more dominant, this will be magnified by the scoring mechanism). Fourth, an expression can be very well recognizable but very unnatural, especially on virtual characters with human faces. Not being able to discriminate between these two measures is a serious problem for the believability of the expression and can negatively impact intended effects on users when the expressions are used (uncanny value effect).

As we are interested in the validity of our method to generate plausible expressions in a wide variety of settings (distances, intensities, blended, viewing angle), we opted for a non-forced choice with additional naturalness measurement. This enables us to study the influence of different conditions on the perception of expressions in detail.

## 2 Three-step FACS-based dynamic affective facial expressions

The development of humanoid virtual characters (aka virtual humans) (Gratch, et al., 2002) involves a number of different steps. Among these steps are multimodal user input, user modeling, virtual agent cognition, emotion and personality, and multimodal output. In this article we focus on a single component of the multimodal output, i.e., generated animation of affective facial expressions.

Our approach is based on facial feature manipulation in three simple steps: feature instrumentation, muscle impact, muscle movement.

First, a set of 18 features (see Table 1) has been instrumented using *3DS Max* on a 3d face based on models from the software package *Facegen*. This set is a subset of the MPEG4 features defined for facial affect expressions (Gratch, et al., 2002). The subset was chosen based on their strong involvement in the generation of the 6 basic expressions of Ekman (Ekman & Friesen, 2003). Each feature is an anchor point attached to a set of vertices of the face. This 3D model "rigging" has to be repeated for each virtual character's morphology. Two rigged example models in *3DS Max* format can be downloaded from http://www.joostbroekens.com or from the humaine website toolbox (http://emotion-research.net/toolbox).

Second, for each feature-expression tuple a muscle pulling vector exists that simply defines the displacement of the feature's anchor point, and thereby indirectly moves the (weighted) vertices (Table 1) (see (Ochs, et al., 2005) section 4.1 for an overview of different approaches towards emotion blending). A vector defines the absolute displacement (in meters) of a feature for an intensity of 1 for each expression. This set of vectors is an important part of the method we propose as it is evaluated in all of the experimental conditions. A new face can reuse this set of vectors, provided that the morphology is comparable (e.g., it would probably not work for emotion expression for animated animals).

Third, we developed a computational model of emotion to represent and express emotional states. The model represents a dynamic emotional state based on factors or categorical emotions or both (this choice is up to developer). The model represents emotion and/or mood and deals with all computational mechanisms needed to represent mood and emotion as well as to integrate affective events over time. The emotion decays to the mood, and the mood is a running average of the emotions over time. Emotions can be triggered by sending affectively labeled events to the model (e.g., [joy, 0.3] to trigger joy with intensity 0.3). The emotional state directly controls the facial expression. Multiple emotions can be triggered at the same time, so the agent can be happy and sad. Emotion decay is as follows. There is an onset, hold (peak) and decay period of 0.4, 0.3 and 0.3 seconds respectively. The intensity of the emotion changes linearly during onset and decay. During hold the intensity is stable at the target intensity of the emotion. Period duration is based on psychometric experiments investigating the timing of forced and spontaneous smiles of people. For example, (Hess & Kleck, 1990) describe an overall total expression time for spontaneous, deliberate joy and disgusted expression equal to 2 seconds, while (K. L. Schmidt, Cohn, & Tian, 2003) report onset peak and offset timings for spontaneous expressions in the order of .5, .2 and .3 respectively (see their Figure 2). These onset timings have recently been replicated (K. Schmidt, Bhattacharya, & Denlinger, 2009). The basis for the onset was 0.5 seconds. To correct for slow muscle speed during the beginning of the onset period (remember, onset looks more like a sinusoid than a straight line) we let the peak duration take a bit from the onset period. Our final onset duration equals 0.4 seconds. Hold (peak) duration differs greatly for spontaneous versus deliberate or social expressions and for different parts of the face. Shorter duration is found for social smiles. Because we also wanted to measure the difficultly of recognition, we have opted for a reduced total expression phase equal to 1 second so that variation would be induced in the number of times the subject chose to see the stimulus before rating it. To keep the total phase to 1 second, we set the hold period to 0.3 seconds. We chose to cut down the hold (peak) period as this is the least influenced by expression dynamics (there is no change in intensity during the hold). As indicated by participants during debriefing, our total phase duration was indeed too short, just as we intended for this study. Our data confirm this, as all of the emotions need more than 1 view on average to decide. None of the participants mentioned unnatural timings, though many mentioned that the total phase was too short. For actual use of the expressions we recommend varying the hold and decay periods so that the total expression takes between 1 and 2 seconds.

For the generation of an expression based on the (possibly blended) emotional state, we use the following simple mechanisms. The final displacement for each feature is based on a simple additive blending model. Each feature's final displacement is the result of the sum over the product the emotion intensities times the feature displacement for the expression prototype of those emotions. If the model is configured to use categorical emotions then the emotional state *is* the vector of emotion intensities. If the model is configured to use factor-based emotions then the emotion intensities are based on the distance to the emotion prototypes in factor space. As such, our model uses geometric intensity (Bartneck & Reichenbach, 2005), i.e., a doubling of emotion intensities results in a doubling of movement of the involved features. If the emotion has decayed, the mood is expressed. Mood is expressed in the same way, but always with a closed mouth (jaw feature). Mood expression has not been evaluated in the experiments.

Our three step approach separates the complexity involved in facial expression modeling by looking at feature selection and weighting, muscle direction and dynamics separately. This is a simple and easy to understand approach, and as we show in the experimental results, it produces good recognition, has the ability to manipulate intensity and the ability to show blended emotions.

## 3   Experimental Setup, Results and Discussion

To evaluate the reliability of our method we investigate the effect of 4 variables that influence perception of the expression: we vary *face morphology*, *viewing angle*, *distance* and *intensity*. We test the following assumptions. First, different facial morphologies have the comparable recognition accuracy when using the same feature vectors and vertex weights (i.e., the method is consistent with respect to different facial morphologies). We tested the difference between two virtual character faces, a female and a male face (Figure 1). These morphologies are different with respect to vertices and texture. Second, lateral expression recognition involves a loss of recognition accuracy, but the perception of positive versus negative should not be influenced. We tested the difference between a lateral 90 degrees view versus a frontal 180 degrees view (Figure 1). Third, higher geometric intensity results in higher perceived intensity for all expressions. We tested two intensity levels, 1.2 versus 2.4 times the weights in Table 1. Fourth, increased distance should not influence recognition accuracy if the resolution is still sufficient to express the emotion. We tested two distances between the user and the virtual character, 1 meter and 3 meters (Figure 1). To test these hypotheses, we showed the same facial expressions in 6 experimental conditions:
1. High intensity male at 1 meter, frontal (used as control for other conditions)
2. High intensity male at 3 meters, frontal.
3. High intensity female at 1 meter, frontal.
4. Low intensity male at 1 meter, frontal.
5. Low intensity male at 1 meter, frontal.
6. High intensity male lateral view

The stimuli consisted of 6 basic emotions (joy, sad, angry, surprised, disgusted, and afraid) and 4 blended emotions (enthusiastic, furious, frustrated and evil). Dynamic facial expressions were generated and presented in the 3d world, with onset, hold and decay timings as mentioned earlier. To appear natural, the virtual character had a default animation for eye-blinking (random periods of about 2 seconds), eye-movement (looking left, straight and right), body, and head movement (slowly wobbling from left to right). Such animations are easy to blend in with the current model of affective expressions. This results in a total of 10 stimuli per condition. Users were presented with one stimulus at the time. They could repeat the stimulus by clicking on the right mouse button. When a subject was ready rating the stimulus they pressed the space bar to see the next stimulus. The experiment took between 15 and 30 minutes to complete.

The set of participants (n=19) included 5 females and 14 males, and was a mix of Western (n=13) and Chinese subjects (n=6), aged between 22 and 36 (mean=27.5, std=3.5). Subjects were recruited through a mailing list, and in person. Subjects could use their own computer, but were specifically asked to only participate using a 15 inch screen with a resolution >=1024x768 positioned in front of the subject at normal distance. All participants viewed all six conditions as well as all 10 expressions (6x10 within subject design), resulting in a total of 60 stimuli to rate per subject.

In terms of outcome measures, we collected the following. For each stimulus participants rated their perception of *intensity on ten emotion labels* on a 5-point Likert scale. The on-screen order of the labels was random between subjects to eliminate sequence effects during rating. Perceived *naturalness* of the expression was rated on a 5-point Likert scale. The number of clicks per stimulus is an objective measure of difficulty of recognition (*nrShows*). For each stimulus we calculated the average perceived intensity across all ten intensity ratings per stimulus (*mean_intensity*), as well as the *target intensity*, which is the participant's perceived intensity of the targeted emotion portrayed by the expression (so for an expression of joy this is the intensity of perceived joy). This means we have a total of 14 outcome variables. All data were analyzed using SPSS 16.

We first discuss the analyses of reliability based on the variation introduced by the four independent variables *face morphology*, *viewing angle*, *distance* and *intensity*. We also look at emotion specific effects of these variables.

Second, we discuss how the individual expressions are perceived (the basic ones: joy, sad, angry, fear, surprised, disgusted, and the blended ones: evil, frustrated, furious and enthusiastic), by analyzing the distribution of the perceived intensities per expression (and a measure of recognition accuracy).

**Table 1**. Feature vectors per expression. Vectors denote [z,x,y] pairs, where positive z, x, and y represent a displacement to the front, to the right and down. Displacement is in meters. Please note the intensity multiplier in the text of the article.

| | surprise | happy | sad | angry | disgust | fear |
|---|---|---|---|---|---|---|
| Jaw | -0.005, 0, 0.01 | 0,0,0 | 0, 0, 0 | 0, 0, 0 | -0.001, 0, 0.001 | -0.002, 0, 0.003 |
| Nostrils | 0, 0, 0 | 0, 0, 0 | 0, 0, 0 | 0, 0, 0 | 0, 0, -0.008 | 0, 0, 0 |
| LipLowerLeft | 0, -0.001, -0.001 | -0.002, 0.001, 0.001 | 0, 0.001, 0.001 | 0, -0.002, 0 | -0.004, 0.002, 0.002 | 0, 0, 0.002 |
| LipLowerRight | 0, 0.001, -0.001 | -0.002, -0.001, 0.001 | 0, -0.001, 0.001 | 0, 0.002, 0 | -0.004, -0.002, 0.0025 | 0, 0, 0.002 |
| LipUpperLeft | 0, -0.002, -0.001 | -0.001, 0.001, -0.001 | 0, 0.001, 0.001 | 0, -0.002, -0.002 | -0.002, 0.002, -0.0045 | 0, 0, -0.002 |
| LipUpperRight | 0, 0.002, -0.001 | -0.001, -0.001, -0.001 | 0, -0.001, 0.001 | 0, 0.002, -0.002 | -0.002, -0.002, -0.0045 | 0, 0, -0.002 |
| LipCornerLeft | 0, -0.001, 0 | -0.005, 0.009, -0.007 | 0, 0.002, 0.007 | 0, -0.004, 0 | 0, -0.001, 0 | 0, 0.002, 0.003 |
| LipCornerRight | 0, 0.001, 0 | -0.005, -0.009, -0.007 | 0, -0.002, 0.007 | 0, 0.004, 0 | 0, 0.001, 0 | 0, -0.002, 0.003 |
| CheekLeft | 0, 0.0, 0.004 | 0, 0, -0.011 | 0, 0, 0 | 0, 0, 0 | 0, 0, -0.003 | 0, 0, 0 |
| CheekRight | 0, 0.0, 0.004 | 0, 0, -0.011 | 0, 0, 0 | 0, 0, 0 | 0, 0, -0.003 | 0, 0, 0 |
| LidLowerLeft | 0, 0, 0.001 | 0, 0, -0.0017 | 0, 0, 0 | 0, 0, 0.001 | 0, 0, -0.0025 | 0, 0, 0.002 |
| LidLowerRight | 0, 0, 0.001 | 0, 0, -0.0017 | 0, 0, 0 | 0, 0, 0.001 | 0, 0, -0.0025 | 0, 0, 0.002 |
| LidUpperLeft | 0, 0, -0.003 | 0, 0, 0.0015 | 0, 0, 0.001 | 0, 0, 0 | 0, 0, 0.002 | 0, 0, -0.003 |
| LidUpperRight | 0, 0, -0.003 | 0, 0, 0.0015 | 0, 0, 0.001 | 0, 0, 0 | 0, 0, 0.002 | 0, 0, -0.003 |
| BrowInnerLeft | 0, 0, -0.005 | 0, 0, 0.0 | 0, 0, -0.005 | 0, -0.013, 0.012 | 0, -0.013, 0.004 | 0, -0.008, -0.006 |
| BrowInnerRight | 0, 0, -0.005 | 0, 0, 0.0 | 0, 0, -0.005 | 0, 0.013, 0.012 | 0, 0.013, 0.004 | 0, 0.008, -0.006 |
| BrowOuterLeft | 0, 0, -0.005 | 0, 0, 0.0 | 0, 0, 0.006 | 0, 0, 0.003 | 0, -0.002, 0 | 0, 0, 0.004 |
| BrowOuterRight | 0, 0, -0.005 | 0, 0, 0.0 | 0, 0, 0.006 | 0, 0, 0.003 | 0, 0.002, 0 | 0, 0, 0.004 |

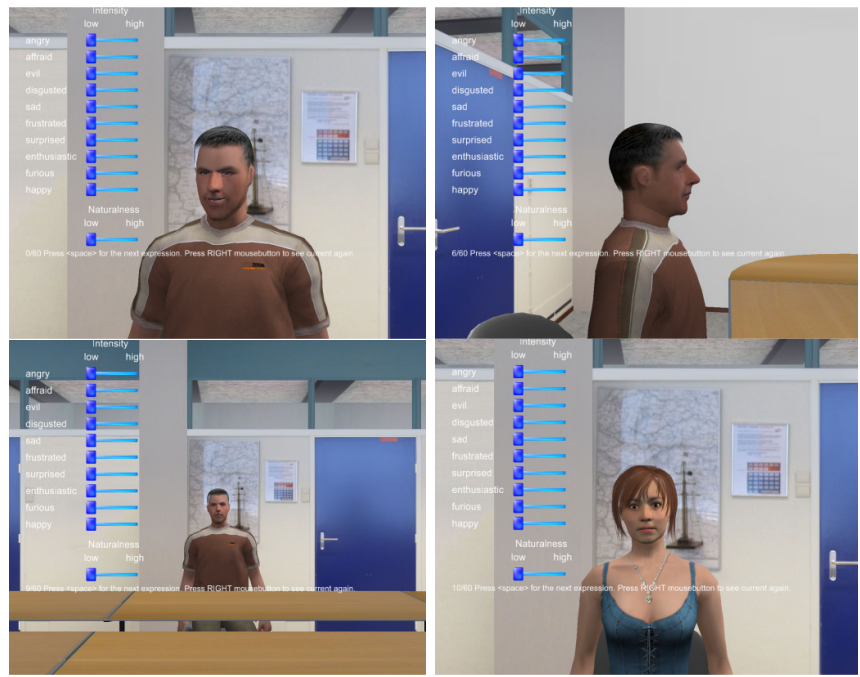

**Fig. 1**. Experiment interface showing four of the six conditions; frontal, lateral, far distance, and female virtual character views.

## 3.1 Manipulations of character type, angle of view, distance and intensity

### 3.1.1 Effect of face morphology.

We did not find evidence for significant influence of morphology on the perception of expressions, as showed by the absence of a significant effect in a multivariate analysis repeated measures ANOVA ($F(13,6)=1.5$, ns) testing the effect of face type.

Even though a main effect is missing, it could be the case that e.g. all female expressions look happier than the male ones, but this effect is not noticeable in the overall effect just reported. Therefore, we also investigate univariate effects[2] of morphology on the average intensity of individual labels. We found main effects of morphology on the perceived intensity of *fear* and *anger* in expressions. On average, subjects perceived more *fear* ($F(1,18)=5.06$, $p<0.05$) in the female expressions (Mean=0.12) than in the male expression (Mean=0.09). Also, the perception of *anger* in female expressions (Mean=0.17) was significantly higher ($F(1,18)=10.1$, $p<0.01$) than the perception of *anger* in male expressions (Mean=0.12).

To investigate if there are expression-specific effects related to these morphology differences, e.g., to find out if the female model looks angrier due to a particular expression, we investigate univariate effects of the interaction between morphology and emotion expression. We found emotion-specific effects for the *naturalness* of the expressions ($F(5.69, 102)=2.91$, $p<0.02$) as well as the perception of *fear* in the expressions ($F(3.22, 58)=2.76$, $p<0.05$). In detail, the expression of *furious* is perceived less natural (paired $t(18)=-5.1$, $p<0.001$) in the female (Mean=0,26) than in the male model (Mean=0.56). Further, the expression of *surprise* was perceived as more fearful (paired $t(18)=2.91$, $p<0.01$) for the female (Mean=0.38) than for the male model (Mean=0.18), explaining why the female looks more fearful. Confusion between fear and surprise is a know phenomenon in psychology (Elfenbein & Ambady, 2002) so we do not consider this a typical problem of our model.

To summarize, the female model looks angrier than the male model, but this is not attributable to a particular expression as we did not find an emotion specific effect. The expression of surprise in the female model looks more fearful than in the male model, and the expression of furious look less natural in the female model. Other than these effects, no influence of morphology was found, as confirmed by the main effect analysis. This indicates two important things:
1. Both morphologies are comparable, and can be taken together in further analysis of the recognition accuracy (Section 3.2).
2. The muscle vectors (Table 1) we have defined for each expression can be used for different character faces, as long as the faces are "rigged" in such a way that the muscle attachment to the 3D vertex model is comparable (step 1). This indicates that our process is repeatable, resulting in a reliable instrumentation.

---

[2] Sometimes the investigation of univariate effects is only done if a main multivariate effect is found. However, as we are specifically interested in the expression details, we do this analysis anyway. For all univariate analysis, we use Greenhouse-Geisser degree of freedom correction for robustness, as our data do not conform to the sphericity assumption. To compare specific means we use paired t-tests.

### 3.1.2 Effect of lateral versus frontal view.

A significant main effect was found (repeated MANOVA, $F(13,6)=13.2$, $p<0.01$) for the viewing angle of the person towards the virtual character. This indicates that angle is an important factor to take into account, as one could hypothesize based on the fact that some facial features will be more difficult to discern if the view towards the avatar is a lateral one (such as inner-eyebrow features). To provide a more detailed picture of what happens exactly, we analyzed univariate main and interaction effects.

With regards to main effects of viewing angle, frontal views (Mean=0.079) are perceived to be more *furious* ($F(1, 18)=5.37$, $p<0.05$) than lateral views (Mean=0.051). Further, frontal views (Mean=0.14) are perceived to be more *sad* ($F(1, 18)=20.0$, $p<0.001$) than lateral views (Mean=0.089). For both the *mean intensity* rating as well as the *target intensity* rating frontal views (Mean=0.089 and Mean=0.446, respectively) are perceived more intense ($F(1, 18)=9.87$, $p<0.01$ and $F(1, 18)=10.7$, $p<0.01$, respectively) than lateral views (Mean=0.079 and Mean=0.363, respectively).

With respect to interaction effects, we found emotion specific effects of viewing angle for the perceived *naturalness* ($F(6,40, 115)=2.37$, $p<0.05$), for the perception of *enthusiasm* in the expressions ($F(2.41, 43)=3.40$, $p<0.05$), for the perception of *furious* in the expressions ($F(3.30, 59)=3.20$, $p<0.05$), and for the perception *sadness* in the expressions ($F(3.42, 62)=4.58$, $p<0.01$). To be more precise about these interaction effects, the expression of *fear* is perceived less natural ($t(18)=2.48$, $p<0.05$) in the lateral view (Mean=0.46) than in the frontal view (Mean=0.61). The expression of *anger* is perceived less natural ($t(18)=2.96$, $p<0.05$) in the lateral (Mean=0.43) than in the frontal view (Mean=0.61). The expression of *enthusiasm* is perceived as less enthusiastic ($t(18)=2.24$, $p<0.05$) in the lateral view (Mean=0.13) than the frontal view (Mean=0.34). The perception of *furious* is perceived as less furious ($t(18)=2.70$, $p<0.05$) in the lateral view (Mean=0.34) compared to the frontal view (Mean=0.54). Finally, the perception of *frustration* is perceived as less sad ($t(18)=3.50$, $p<0.01$) in the lateral view (Mean=0.20) compared to the frontal view (Mean=0.46).

To summarize, these data show that lateral views are perceived to be less intense, regardless the emotion presented. Also, two emotions are perceived to be less natural: fear and anger. Further, there are some specific differences between lateral and frontal views that point towards two important consequences for presenting a lateral view to the user. These consequences are:
1. a loss of subtlety, combined with
2. a tendency to rate expressions as basic emotions instead of blended ones.

Evidence for these consequences is found in the following three observations. First, lateral expression of furious is perceived as less furious than frontal expressions (see above), but the perceived angry component stays the same (Mean=0.46), indicating that the basic emotion component is still perceived in the blend while the more specific one is not. Second, the same pattern can be found for enthusiastic. Enthusiastic expressions are perceived less enthusiastic but the perceived joy component is not significantly different (Means 0.33 in the lateral and 0.44 in the frontal view). Third, expressions of frustration are perceived to be less sad in the lateral view, while this was the dominant interpretation for it in the frontal view. This indicates that frustration loses its sadness component, which is probably due to the loss of detail with regards to the difference between inner-eye brow and out eye-brow as well as mouth curvature. These two features are hardly visible in lateral view. These observations indicate a loss of subtlety combined with a focus on basic emotions. This is consistent with work of others (Arellano, et al., 2008) who also report that blended emotions are often perceived as extreme (basic) ones. These two consequences are important to keep in mind when modeling emotions for virtual characters in 3d worlds, in particular when there is the need to express subtle emotions (e.g., because one wants to express empathic emotions) but the user/virtual character has the freedom to move around.

### 3.1.3 Effect of distance and intensity.

A significant main effect was found for intensity (repeated MANOVA, $F(13,6)=13.2$, $p<0.02$). No main effect was found for distance ($F(13,6)=1.53$, ns), nor for an interaction between distance and intensity ($F(13,6)=0.98$, ns). We first investigate specific effects of distance. Intensity effects are discussed later in this section.

Looking in more detail at possible univariate effects of distance we found a significant effect for *frustration* ($F(1, 18)=5.50$, $p<0.05$). At large distance expressions are perceived to be less frustrating (Mean=0.047) than at close distance (Mean, 0.064). Further we see a significant effect on *mean intensity* as well as *target intensity*. Perceived mean intensity is smaller ($F(1, 18)=11.1$, $p<0.01$) at large distance (Mean=0.073) than at closer distance (Mean=0.079), a difference of about 8%. Target intensity is smaller ($F(1,18)=4.45$, $p<0.05$) at large distance (0.34) than at close distance (0.37), a difference of about 9%. Compared to the effect of manipulated intensity, however, these effects on perceived mean and target intensity are small (e.g., the increase in target intensity due to intensity manipulation is 54%, see below).

Looking at univariate effects of the interaction between distance and intensity, we see only an effect on the perception of *sadness* in the expressions ($F(1,18)=6.75$, $p<0.05$) (Figure 2). It seems that the perception of sadness in expressions at a close distance does not react to intensity manipulation. This could indicate that the low intensity expression of sadness is already quite intense at close distance indicating that the interaction effect we found has something to do with intensity maxing out rather than distance. We will come back to this when discussing the univariate interaction effects between emotion and intensity later in this section.

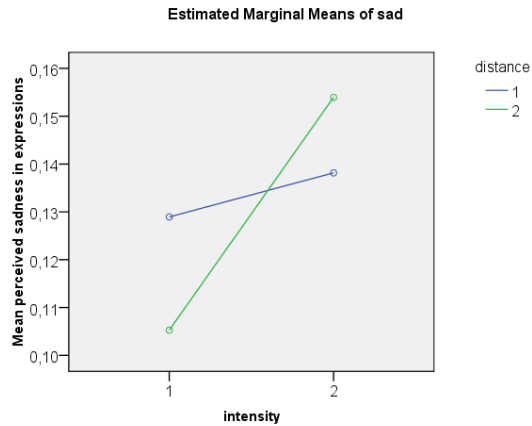

**Fig 2**. Interaction effect between distance and intensity on the amount of sadness perceived in all expressions (the sadness measure).

As neither distance nor the interaction between distance and intensity were significant factors influencing how people perceived the expressions, we now turn to a more detailed analysis of the effects of intensity and the interaction effects between intensity and emotion.

The intensity manipulation was successful and consistent, by which we mean that it influenced the perceived intensity of all emotions in the right direction (i.e., larger geometric intensity resulted in higher perceived intensity). We now discuss our main findings.

No significant univariate main effects were found for intensity on *nrShows* ($F(1, 18)=3.28$, ns) and *naturalness* ($F(1,18)=2.04$, ns), indicating that, in general, intensity did not influence the difficulty of recognizing the emotion nor the naturalness of the expressed emotions. However the perceived *naturalness* was influenced by the interaction between emotion and intensity ($F(5.22, 93.9)=2.92$, $p<0.05$), indicating that the naturalness of some expressions were affected by intensity in specific ways.

In detail we found that *enthusiastic* was perceived less ($t(18)=2.52$, $p<0.05$) natural in the intense condition (Mean=0.28) than in the less intense condition (Mean=0.41). We found that *furious* was perceived more ($t(18)=-2.22$, $p<0.05$) natural in the intense condition (Mean=0.49) than in the less intense condition (Mean=0.39). Finally we found that *joy* was perceived less ($t(18)=2.57$, $p<0.05$) natural in the intense condition (Mean=0.36) than in the less intense condition (Mean=0.49). Other emotion specific effects were not significant, meaning that the effect of intensity on the naturalness of the expressions is limited to three emotions; joy, furious and enthusiastic.

We now analyze if intensity manipulated the perceived intensity of the target emotion. We see that perceived *target intensity* is higher ($F(1,18)=92.6$, $p<0.001$) in the high intensity condition (Mean=0.43) than in the low intensity condition (Mean=0.28). This means that overall the target intensity increased by about 54%. This is consistent with previous work by (Bartneck & Reichenbach, 2005), who also report successful manipulation of perceived intensity using geometric intensity. Emotion expression specific effects of manipulated intensity on perceived *target intensity* are shown in Table 2 and Figure 3. These results show that for half of the expression the manipulation was significant while for the other half it was not.

Lack of significant influence on perceived intensity can be because (a) the high intensity expression was not strong enough, (b) the low intensity expression was too strong, or (c) because there is a problem with the recognition of the expression itself. For *enthusiastic* it seems the manipulation did not affect the perception of intensity because of explanation *c*. As mentioned above, high enthusiastic expression has a far lower naturalness score. This means that high intensity enthusiasm itself has recognition problems. For *evil* it seems that explanation *a* would be most fitting. High intensity evil is not strong enough, but evil itself is well recognized (see later). For the expression of *frustration* it seems that it is not recognized (explanation *c*), as shown by the low scores for both the low and the high intensity conditions. As our analysis of the perception of individual emotion expression (next section) will show, frustration was indeed one of the emotions that was not clearly identified. This also explains the non-response to the intensity manipulation. For *joy* and *sad* the explanation seems to be a combination of *a* and *b*. The low intensity emotion is quite high on intensity compared to the intensity of the other expressions, while the high intensity does not add much to it. This explanation is backed up by our *distance X intensity* interaction effect for sadness mentioned earlier.

We now address the strong effect of intensity on the expression of furious. This can be explained by the fact that low intensity furious is something that is conceptually inconsistent (how can you be a little furious?). This is reflected by the intensity rating: high intensity furious is seen as furious (and, as mentioned above, more natural than low intensity furious), low intensity furious is seen as angry (see Figure 4, and detailed analysis of furious in the next section).

To give some more insight in what happens with individual expressions, we present the details of the perceived anger intensity in the different expressions (Figure 4). We see that expressions related to anger (anger, evil, frustration, furious) are perceived as more angry. We found the same pattern for sad (and frustrated), and joy (and enthusiastic). This means that the basic emotion composing the blended expression still influence how people perceive this blended emotion.

We also checked if we could replicate the result by (Bartneck, 2001) who found that perceived intensity correlates with perceived naturalness. Indeed we found a significant correlation (Pearson's r=0.50, n=19, p<0.05) between a subject's overall perception of naturalness and overall rating of target intensity. This means that people who judge the expressions as natural also judge them as intense (please note that this is not the same as concluding that *expressions* that are rated high on naturalness are also rated high on intensity, it only tells us something about people in general).

To summarize, distance manipulation was not significant, apart from several isolated effects. Very large distances would influence expression perception but this is a different question that relates to screen and eye resolution loss. Intensity manipulation was quite successful, especially given the simple geometric operation used. Finally, people who rate high on intensity also rate high on naturalness.

**Tabel 1.** Intensity manipulation results for individual emotions (significant in **bold**).

|              | Low intensity | High intensity | paired t(18) | Sig. (2-tailed) |
|--------------|---------------|----------------|--------------|-----------------|
| afraid       | 0.23          | 0.34           | -2.10        | **0.05**        |
| angry        | 0.28          | 0.43           | -2.37        | **0.03**        |
| disgusted    | 0.18          | 0.34           | -2.76        | **0.01**        |
| enthusiastic | 0.18          | 0.23           | -0.78        | 0.45            |
| evil         | 0.20          | 0.29           | -1.22        | 0.24            |
| frustrated   | 0.04          | 0.10           | -1.37        | 0.19            |
| furious      | 0.05          | 0.32           | -3.47        | **0.00**        |
| joy          | 0.28          | 0.36           | -1.84        | 0.08            |
| sad          | 0.33          | 0.38           | -1.22        | 0.24            |
| surprised    | 0.31          | 0.45           | -2.85        | **0.01**        |

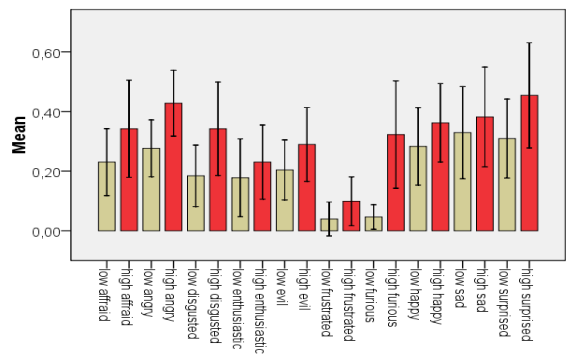

**Fig 3.** Mean (n=19) and 95% confidence intervals for *target intensity* for the low (gray) and high (red) intensity perception of individual expressions.

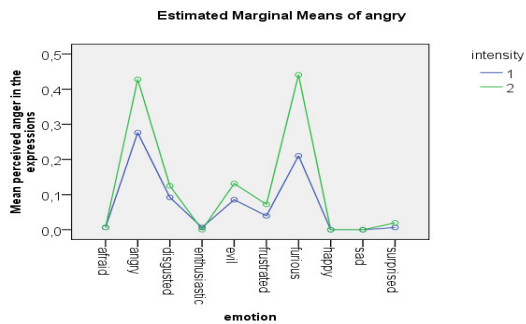

**Fig 4.** Typical effect of intensity manipulation, in this case perceived anger in all expressions.

### 3.2 Perception of individual emotional expressions.

We now analyze the validity of the generated expressions, i.e., what is the affective profile of each expression. For this, we take the three experimental conditions that did not show significant main effects together (male close, female close, male far). Ratings for expressions were averaged across these conditions.

As we are primarily interested in specific effects on the perception of expressions, we performed an univariate analysis. It showed that all of the label ratings, the naturalness and the derived intensity outcome variables were significantly influenced by emotion expression (Table 3 shows details). The difficulty, as measured by the amount of clicks needed to rate the emotion, was not influenced by the emotion, meaning that we can consider the expressions equal in difficulty to recognize. The naturalness of the expression was influenced by emotion, showing that some emotions are more natural than others. Further, mean and target intensity are influenced by emotion, showing that not all of the emotions are perceived with equal intensity.

**Table 3.** Effects (Greenhouse-Geisser) of emotion expressions on peoples rating behavior.

| Outcome | df | error df | F | Sig. |
|---|---|---|---|---|
| nrShows | 4.654 | 83.77017 | 1.554 | 0.186 |
| naturalness | 5.710 | 102.7721 | 7.911 | 0.000 |
| affraid | 2.563 | 46.1409 | 28.591 | 0.000 |
| angry | 3.212 | 57.81257 | 23.703 | 0.000 |
| disgusted | 2.888 | 51.99291 | 13.313 | 0.000 |
| enthusiastic | 1.626 | 29.27163 | 16.863 | 0.000 |
| evil | 3.382 | 60.87729 | 26.260 | 0.000 |
| frustrated | 4.011 | 72.20641 | 3.359 | 0.014 |
| furious | 2.840 | 51.12813 | 16.004 | 0.000 |
| joy | 2.328 | 41.89887 | 42.038 | 0.000 |
| sad | 3.293 | 59.27732 | 43.054 | 0.000 |
| surprised | 2.541 | 45.73829 | 36.891 | 0.000 |
| mean_intens | 3.919 | 70.54259 | 7.549 | 0.000 |
| target_intens | 4.808 | 86.55246 | 6.006 | 0.000 |

Looking first at perceived naturalness (Figure 5), we see immediately that the expression of enthusiasm is perceived far less natural than the other expressions (significant for all t(18)<-2.1, p<0.05). This confirms our earlier analysis that enthusiasm as a blend of surprise and joy probably is too extreme in the high intensity version. Looking at target intensity (Figure 6), we see that frustration has a far lower perceived intensity than the other expressions (significant with t(18)<2.3, p<0.05, except for the difference between frustration and enthusiasm, *ns*). Combined with the fact that the average intensity of frustration is comparable to the other expressions, this means that there is a problem with the recognition of frustration (as confirmed in detailed bar chart for frustration, Figure 7f).

As a measure for quality (recognition accuracy) we propose to look at target_intensity/(average_intensity*nr_of_emotions). This value simply expresses the confusion in a particular expression as the proportion of the target emotion's intensity relative to all perceived emotion intensities for a particular expression. We can clearly see that frustration, enthusiasm and furious are the three worst ones, which is interesting as these are also the automatically generated blended emotions (Table 4). Apparently, these blended emotions are either difficult to interpret, difficult to model, or simply not the linear combination of two basic emotions. As we will see later in the expression-specific rating profiles, these emotions are not recognized consistently as a unique blended emotion, but are perceived as their dominant basic emotion components, explaining their low quality score. This means the affective content is recognized but the label to rate it is not. Disgust also does not seem to score well, however as we will see later, disgust is uniquely recognized, and only has a small confusion with anger (a known confusion with disgust expression in virtual agents). Interestingly, evil scores good, and in any case much better than the other blends. This shows that a blend of joy and anger generates an expression that is uniquely identified as evil. This is an interesting finding and compatible with the emotion and expression of *schadenfreude* (gloating). The other expressions all have high quality and are uniquely identified (Figure 7).

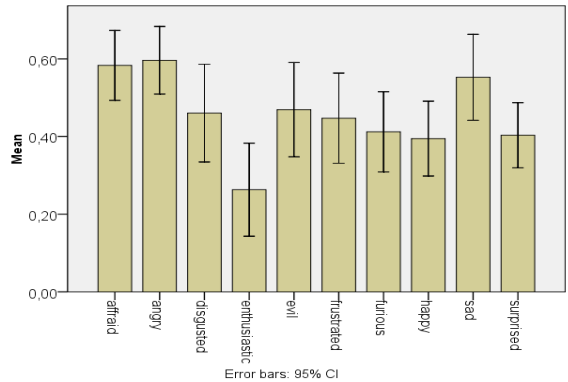

**Fig. 5.** Perceived naturalness of the different emotions.

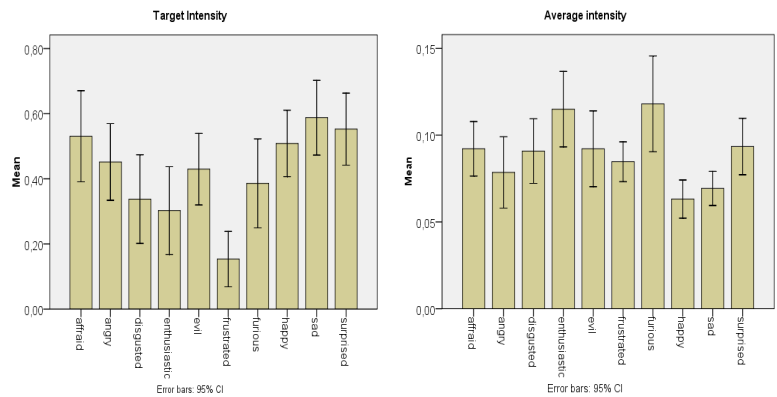

**Fig 6**. Perceived target intensity (left) and average intensity (right) for the different emotions.

**Table 4**. Target and average intensities and recognition coefficient as a measure of accuracy.

|              | Average | Target | Quality |
|--------------|---------|--------|---------|
| affraid      | 0.092   | 0.531  | 0.576   |
| angry        | 0.079   | 0.452  | 0.575   |
| disgusted    | 0.091   | 0.338  | 0.372   |
| enthusiastic | 0.115   | 0.303  | 0.263   |
| evil         | 0.092   | 0.430  | 0.467   |
| frustrated   | 0.085   | 0.154  | 0.181   |
| furious      | 0.118   | 0.386  | 0.327   |
| joy          | 0.063   | 0.509  | 0.806   |
| sad          | 0.069   | 0.588  | 0.848   |
| surprised    | 0.093   | 0.553  | 0.592   |

We now look in detail at how individual expressions have been perceived (Figure 7a-j). We discuss the scoring pattern of each of the ten expressions.

First, the expression of fear is recognized uniquely, and has no confusion with the other labels (confirmed by a paired $t(18)>2.78$ $p<0.01$ for all differences between afraid and other labels). The overall recognition coefficient for afraid is 0.576, meaning that for expressions of fear 57.6% of the perceived intensity is indeed fear.

Second, the expression of anger is uniquely recognized ($t(18)<4.71$, $p<0.001$ for all comparisons), and, like fear, has little confusion, a recognition coefficient of 57.5%.

Third, For the perception of the disgusted expressions, we observe a small confusion with anger ($t(18)=1.91$, $p=0.073$), but otherwise there is no confusion. Disgust does have a rather low recognition coefficient. This can be explained by the fact that disgust is confused with anger, and there are three labels that all have anger components (angry, furious and frustrated). These labels indeed all score relatively high for the disgust expression. Others also found confusion effects for disgust (Arellano, et al., 2008; Noel, et al., 2006 ). In our study the perception of disgust in disgusted expressions was 2.5 times higher than the perception of anger, and the difference was almost significant. This is good result. A change that can be implemented to enhance the difference is an increase in the asymmetry in the expression.

Fourth, the expression of the blended emotion enthusiasm is not uniquely identified as enthusiastic, and also scores high on joy and surprised which are its basic constituents (the difference between joy, surprised and enthusiastic is not significant). Perceived happiness is indeed the most important recognized intensity. The difference between perceived happiness and all other emotions (except enthusiastic and surprised) is significant (all $t(18)>3.77$, $p<0.001$). This means that an important part of the expression of enthusiasm is interpreted as the basic components it is composed of, and not as a separate emotion. This interpretation is supported by our previous findings: the naturalness of high intensity enthusiasm is low indicating that people probably also had difficulty interpreting it, the loss of enthusiasm recognition and shift to basic emotion interpretation in the lateral view indicating that the details of the expression of enthusiasm are quickly lost, and the absence of intensity manipulation combined with a high rating on happiness indicating that enthusiasm is more easily seen as extreme happiness. This makes us believe we should interpret the expressions of enthusiasm as a high intensity variant of joy, and that adding a little bit of surprise to the expression of happiness can make the intensity modulation more effective for the expression of joy. This shows the potential of blending emotions with our method.

Fifth, the expression of the blended emotion evil is uniquely recognized as an expression of evil (all $t(18)>3.01$, $p<0.01$). This is a potentially important finding that we discuss in the next section in more detail.

Sixth, the expression of frustration is not recognized. In fact, it is uniquely identified as sadness ($t(18)>4.21$, $p<0.001$), indicating that the sadness component was dominant even though there was exactly the same amount of sadness as anger present in the expression. This can be due to a combination of the following things. The label frustration was wrongly chosen, the expression of frustration is not a combination of sadness and anger, or the expression of frustration is difficult to recognize. We think there is evidence for all three causes. First, the label was too generic, as shown by the fact that 5 out of ten expressions (Figure 8) have a moderate amount of perceived frustration, all not significantly different from the expression of frustration itself. So, apparently people saw a bit of frustration in many expressions. Second, frustration itself is seen as sadness, and the lateral view condition supports this as in this condition we see a shift to perceiving frustration as sadness. Therefore, our expression of frustration resembled sadness too much. Third, from a theoretical point of view, frustration is not so much an emotional expression but more so an affective state that is rather undifferentiated, adding to the difficult for people to recognize this. These reasons make us believe that, without additional contextual information, frustration, and more general, a combination of anger and sadness on the face of a virtual character is difficult to identify for people.

Seventh, furious has the same pattern as enthusiastic. Here, it is not uniquely identified and anger is its major perceived component ($t(18)>4.74$, $p<0.001$, except the difference between perceived anger and furious, *ns*). This again means that furious is perceived as its basic emotion, even though it is a blend of surprise and anger. This interpretation is supported by our earlier findings that show that high intensity furious expressions are perceived to be very angry, that high intensity furious is more natural than low intensity furious (so furious is a high-intensity emotion per se not a basic one that can also have a low intensity variant), and that the lateral view on furious expressions shifts to being perceived as angry, not furious (meaning that the loss of detail makes people interpret the emotion more in line with its basic constituent). This makes us believe we should interpret furious as a high intensity variant of anger, and that adding a little bit of surprise to anger can make the intensity modulation more effective for anger expressions. Again, this shows the potential of blending emotions with our method.

Eight, the expression of happiness is recognized very well (all $t(18)>6.74$, $p<0.001$). It has a high measure of selectivity (0.81). Analysis of the objective difficulty (*nrShows*) also showed that happiness is easier to recognize than all other expressions (Figure 9), but this was only significant for the difference between joy on one side and angry, frustrated and furious on the other.

Ninth, the expression of sadness is recognized very well too (all $t(18)>7.81$, $p<0.001$). It has a high measure of selectivity (0.85). Again, objective difficulty also shows that sadness is easy to recognize than most other emotions (Figure 9), although this was significant for only two emotions, i.e., the difference between sadness on one side and anger and frustrated on the other.

Tenth, the expression of surprise is uniquely identified (all t(18)>3.93), meaning that also surprise is well recognized as surprise.

To summarize, the basic emotions are all uniquely identified with relatively little confusion, except for disgust which is slightly confused with anger. As this is a result based on three different presentation conditions, we feel confident our method is a reliable and valid one. Our results also nicely replicate work of others with respect to the difficulty to generate disgust (Arellano, et al., 2008). It is also compatible with emotion expression in psychology. Fear expressions often have surprise as second component and sadness, happiness and anger are often the best recognized (Elfenbein & Ambady, 2002). With regards to blended emotions, we found out that a linear combination of basic emotions is in principle able to generate blends. Sometimes this results in a different uniquely recognizable emotion, (as shown by the expression of evil being a blend of joy and angry. Sometimes this results in a tendency to be interpreted as extreme forms of the basic constituent (furious and enthusiastic). It is interesting to note that this indicates that when blending other basic emotions with the expression of surprise, it seems that we can in some case add intensity (or arousal mediating intensity) to the expression.

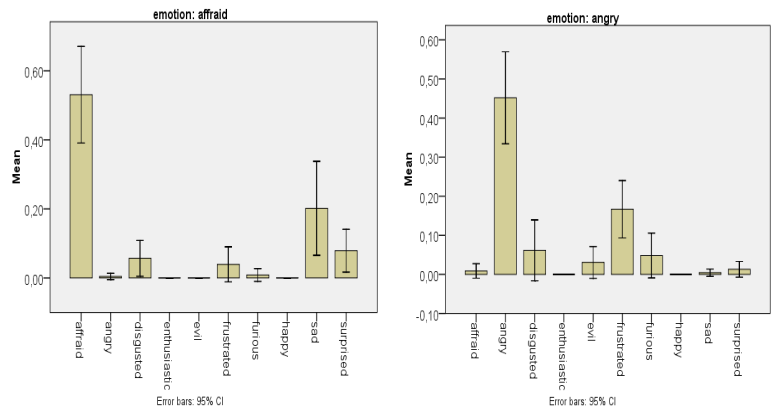

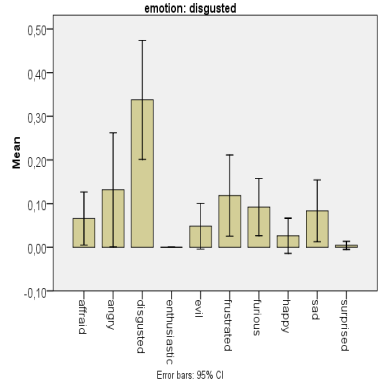
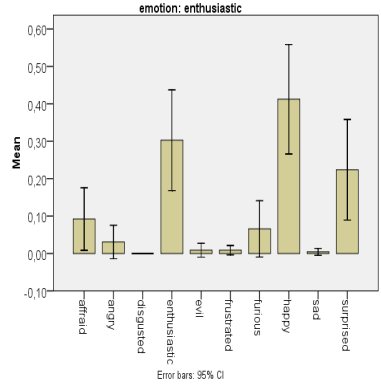
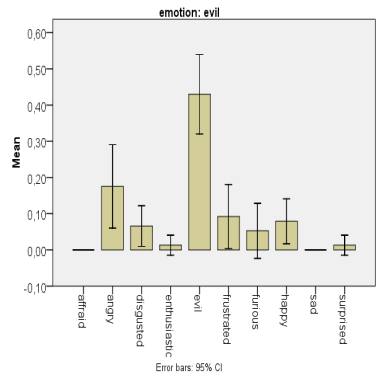
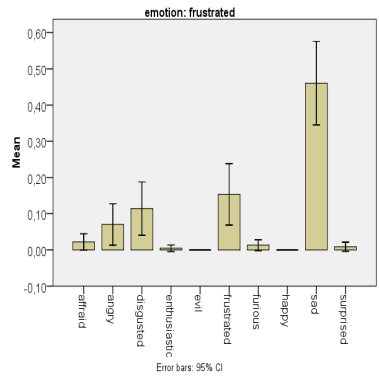
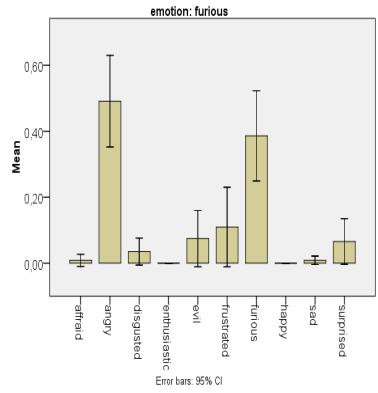
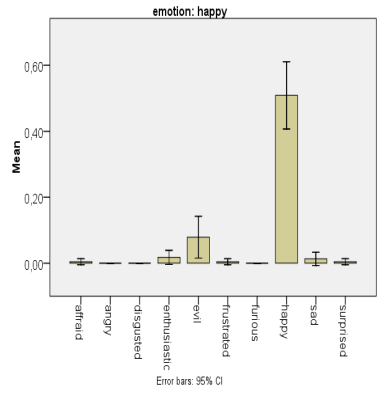

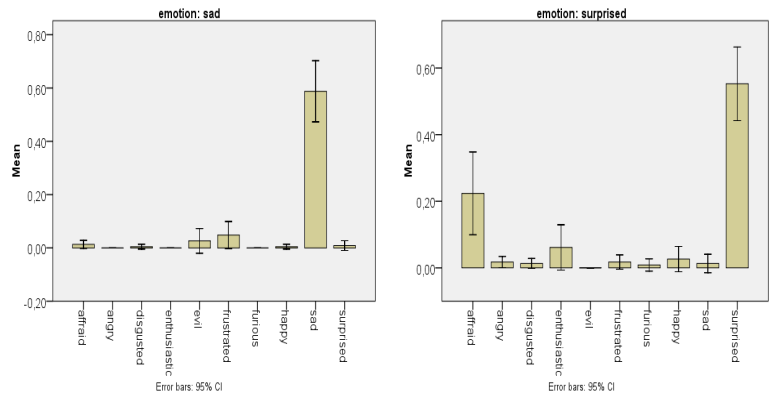

**Fig 7a-j**. Scoring pattern for each of the different expressions. For example, the expression of surprise was recognized as surprise (subjects rated the expression of surprise high on the 5-point Likert scale labeled "surprise").

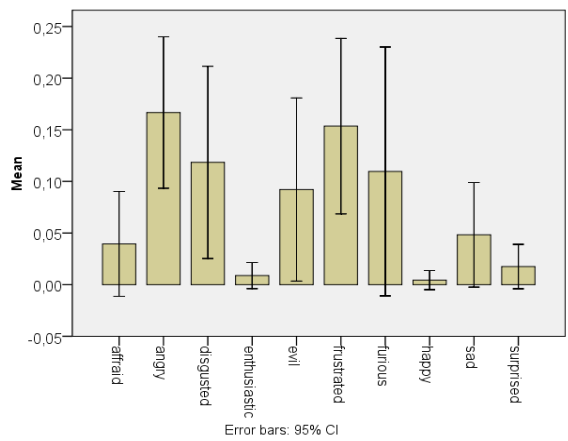

**Fig. 8**. The perception of frustration in the different expressions. For example, angry expressions have a relatively high element of frustration.

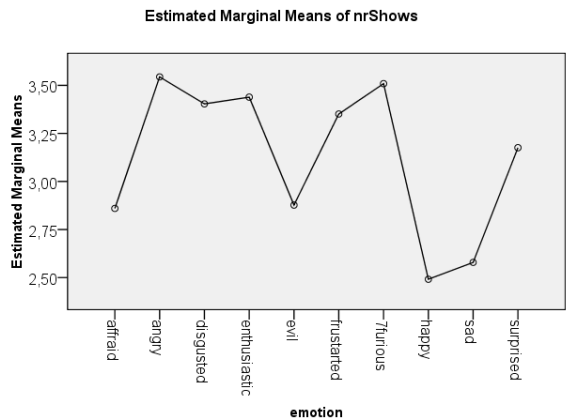

**Fig 9**. Objective difficulty measured by the number of times subjects chose to see the expression before moving to the next one (nrShows).

### 3.3 Evil; blending joy and anger results in a unique expression.

We now turn to an important finding in our study, the perception of evil[3] in the expression of a blended emotion composed of happiness and anger. Evil was composed as a 50/50 linear blend of anger and happiness, without additional validation or modeling efforts. As such, evil in our case is the emergent result of these two basic universal expressions. Interestingly, evil is recognized as evil, not as its basic emotion constituents (as is the case with furious, enthusiasm and frustration). Also, there is virtually no perception of evil in the other expressions, making evil also a very selective label (see Figure 10). To our knowledge this is the first report on the perception of the expression of evil in virtual characters as a uniquely identifiable emotion. As evil is a well-known expression in animation pictures and movies, and is of interest in any setting where aggression needs to be simulated, we feel this is an important contribution for the field of virtual characters.

It would go too far to go into the psychological ramifications of this finding in this article, given that we have tested the expression only with virtual characters. However, we do want to informally sketch some of them. First, the appraisal of evil is related to joy and anger: joy about going to do something you like, angry as empathic reaction towards the one/ones you are doing the thing too. This indicates that the blend could make sense from an appraisal point of view. Second, there is a specific (domain independent) action tendency associated with evil (a more neutral term for evil would be *naughty*, see footnote): going to do something of which you know others disapprove. Third, there is developmental benefit for being able to express and recognize the expression. In a developmental context it is important to know if a child knows the rules but chooses to ignore them (your reaction would be punishment/anger), or does not know the rules and is playing incidentally against them (your reaction would be to explain why the action is dangerous or unwanted). Fourth, our participants were Chinese and Western subjects, indicating that the perception of evil is probably not culture specific. Evidence for this can also be seen in the expression being the same in Asian as well as Western cartoons (angry eyes and smiling). Fifth, the expression of evil and shadenfreude (gloating) resemble each other, but evil is a different emotion related to an anticipation and action tendency, while schadenfreude (gloating) is an evaluation of a (usually past) situation. We are currently studying the perception of evil in a wider set of stimuli, as well as identifying psychological implications.

---

[3] Note that related terms for what we mean with evil include naughty, disobedient and mean.

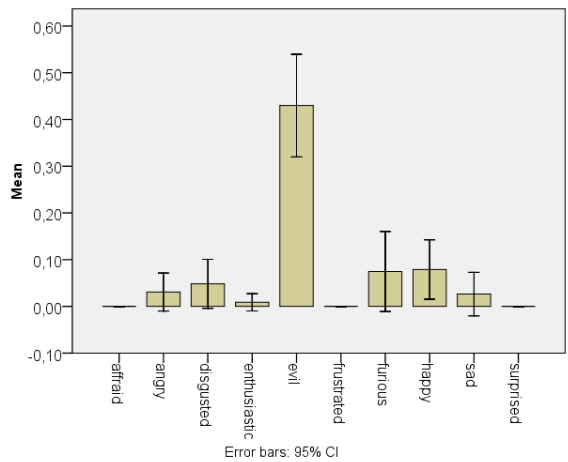

**Fig. 10**. The perception of evil in the different expressions. For example, surprised has no element of frustration (no error bar, means no deviation from 0, means none of the subjects feels that expression of surprise has a component of evil in it).

## 4   Conclusion

We have presented an easy to use method for the generation of dynamic affective facial expression. By now we have used our method in several studies (Broekens et al., 2012; Ham & Broekens, 2011) and it is in active use by a commercial company (CleVR.net) specialized in VR training and treatment. It is freely downloadable from http://www.joostbroekens.com and is available at the *Humaine* website in the toolbox section in the category of emotional expression (http://emotion-research.net/toolbox). We have experimentally addressed the reliability of the method by addressing four key factors that could influence the user's perception of affective facial expressions in virtual characters. We addressed validity of the expressions by investigating the accuracy and expression profiles for 6 basic expressions and 4 blended expressions.

The effect of face morphology and the effect of distance on perception were not significant. We found evidence that a lateral presentation involves a loss of subtlety combined with a focus on perceiving blended expressions as basic ones. These findings should be taken into account when modeling emotions for virtual characters in 3d worlds, in particular when there is the need to express subtle emotions and when the user or virtual character can move around. A lateral view might reduce the intended effect of subtle expression differences. We found that geometric intensity manipulation had the intended effect. For half of the expressions the effect was not significant, but plausible explanations were provided for the lack of significance.

With regards to the facial affective expression recognition accuracy itself, we found that all basic emotions except the expression of disgust were uniquely identifiable. Disgust was slightly confused with anger, but this is a known difficulty in literature (Arellano, et al., 2008; Bartneck & Reichenbach, 2005).

Further, our approach enables the blending of emotions. Two of the four tested blends showed a user perception that is consistent with how they were constructed. One blend showed to be confusing. The fourth blend, *evil* a blend of joy and anger, was uniquely identified by subjects. This alone is an interesting result and deserves further study. It could point towards the universal recognition of evil (or naughty or any other of the related labels). The fact that also blended emotions produced consistent recognition patterns gives additional support to the reliability of our method as well as the validity of the expressions that result from it.

## References

Arellano, D., Varona, J., & Perales, F. J. (2008). Generation and visualization of emotional states in virtual characters. *Computer Animation and Virtual Worlds, 19*(3-4), 259-270.

Bartneck, C. (2001). How Convincing is Mr. Data's Smile: Affective Expressions of Machines. *User Modeling and User-Adapted Interaction, 11*(4), 279-295.

Bartneck, C., & Reichenbach, J. (2005). Subtle emotional expressions of synthetic characters. *International Journal of Human-Computer Studies, 62*(2), 179-192.

Bevacqua, E., Mancini, M., Niewiadomski, R., & Pelachaud, C. (2007). An expressive ECA showing complex emotions *Proceedings of the AISB Annual Convention* (pp. 208-216). Newcastle, UK: AISB.

Breazeal, C., & Scassellati, B. (2000). Infant-like Social Interactions between a Robot and a Human Caregiver. *Adaptive Behavior, 8*(1), 49-74.

Broekens, J. (2010). Modelling the experience of emotion. *International Journal of Synthetic Emotions, 1*(1), 1-17.

Broekens, J., & DeGroot, D. (2004). Scalable and Flexibel Appraisal Models for Virtual Agents. In Q. Mehdi & N. Gough (Eds.), *Proceedings of the Fifth Game-on International Conference* (pp. 208-215).

Broekens, J., Harbers, M., Brinkman, W.-P., Jonker, C., Bosch, K., & Meyer, J.-J. (2012). Virtual Reality Negotiation Training Increases Negotiation Knowledge and Skill. In Y. Nakano, M. Neff, A. Paiva & M. Walker (Eds.), *Intelligent Virtual Agents* (Vol. 7502, pp. 218-230): Springer Berlin Heidelberg.

Buisine, S., Abrilian, S., Niewiadomski, R., Martin, J.-C., Devillers, L., & Pelachaud, C. (2006). Perception of Blended Emotions: From Video Corpus to Expressive Agent. In

J. Gratch, M. Young, R. Aylett, D. Ballin & P. Olivier (Eds.), *Intelligent Virtual Agents* (Vol. 4133, pp. 93-106): Springer Berlin / Heidelberg.

Carolis, B. d., Pelachaud, C., Poggi, I., & Steedman, M. (2004). APML, a Mark-up Language for Believable Behavior Generation. In H. Prendinger & M. Ishizuka (Eds.), *Life-like characters: tools, affective functions, and applications* (pp. 65-85): Springer.

Core, M., Traum, D., Lane, H. C., Swartout, W., Gratch, J., van Lent, M., et al. (2006). Teaching Negotiation Skills through Practice and Reflection with Virtual Humans. *SIMULATION, 82*(11), 685-701.

Ekman, P., & Friesen, W. (2003). *Unmasking the face: A guide to recognizing emotions from facial expressions.*: Cambridge, MA: Malor Books.

Elfenbein, H. A., & Ambady, N. (2002). On the universality and cultural specificity of emotion recognition: a meta-analysis. *Psychological bulletin, 128*(2), 203.

Gilleade, K. M., & Dix, A. (2004). Using frustration in the design of adaptive videogames *Proceedings of the 2004 ACM SIGCHI International Conference on Advances in computer entertainment technology*. Singapore: ACM.

Graesser, A. C., Chipman, P., Haynes, B. C., & Olney, A. (2005). AutoTutor: an intelligent tutoring system with mixed-initiative dialogue. *Education, IEEE Transactions on, 48*(4), 612-618.

Gratch, J., & Marsella, S. (2001). Tears and fears: modeling emotions and emotional behaviors in synthetic agents *Proceedings of the fifth international conference on Autonomous agents* (pp. 278-285). Montreal, Quebec, Canada: ACM.

Gratch, J., Marsella, S., & Petta, P. (2009). Modeling the cognitive antecedents and consequences of emotion. *Cognitive Systems Research, 10*(1), 1-5.

Gratch, J., Rickel, J., Andr\, E., \#233, Cassell, J., Petajan, E., et al. (2002). Creating Interactive Virtual Humans: Some Assembly Required. *IEEE Intelligent Systems, 17*(4), 54-63.

Hall, L., Woods, S., Aylett, R., Newall, L., & Paiva, A. (2005). Achieving Empathic Engagement Through Affective Interaction with Synthetic Characters *Affective Computing and Intelligent Interaction* (pp. 731-738).

Ham, W. v. d., & Broekens, J. (2011). *The Effect of Dominance Manipulation on the Perception and Believability of an Emotional Expression.* Paper presented at the Workshop on Standards in Emotion Modelling, Leiden.

Hess, U., & Kleck, R. E. (1990). Differentiating emotion elicited and deliberate emotional facial expressions. *European Journal of Social Psychology, 20*(5), 369-385.

Heylen, D., Nijholt, A., Akker, R. o. d., & Vissers, M. (2003). Socially intelligent tutor agents. In R. Aylett, D. Ballin & T. Rist (Eds.), *Proceedings of Intelligent Virtual Agents (IVA)* (pp. 341-347).

Heylen, D., Vissers, M., op den Akker, R., & Nijholt, A. (2004). Affective Feedback in a Tutoring System for Procedural Tasks *Affective Dialogue Systems* (pp. 244-253).

Hudlicka, E. (2003). To feel or not to feel: The role of affect in human-computer interaction. *International Journal of Human-Computer Studies, 59*(1-2), 1-32.

Hudlicka, E. (2008). Affective Computing for Game Design *Proceedings of the 4th Intl. North American Conference on Intelligent Games and Simulation* (pp. 5-12).

Hudlicka, E., & Broekens, J. (2009). Foundations for modelling emotions in game characters: Modelling emotion effects on cognition *Affective Computing and Intelligent Interaction and Workshops, 2009. ACII 2009. 3rd International Conference on* (pp. 1-6).

IJsselsteijn, W. A., Ridder, H. d., Freeman, J., & Avons, S. E. (2000). Presence: concept, determinants, and measurement *Proceedings of SPIE 3959* (pp. 520-529).

Kätsyri, J., & Sams, M. (2008). The effect of dynamics on identifying basic emotions from synthetic and natural faces. *International Journal of Human-Computer Studies, 66*(4), 233-242.

Lance, B. J., & Marsella, S. C. (2008). A model of gaze for the purpose of emotional expression in virtual embodied agents *Proceedings of the 7th international joint conference on Autonomous agents and multiagent systems - Volume 1* (pp. 199-206). Estoril, Portugal: International Foundation for Autonomous Agents and Multiagent Systems.
McQuiggan, S. W., & Lester, J. C. (2007). Modeling and evaluating empathy in embodied companion agents. *International Journal of Human-Computer Studies, 65*(4), 348-360.
Niewiadomski, R., & Pelachaud, C. (2007). Fuzzy Similarity of Facial Expressions of Embodied Agents. In C. Pelachaud, J.-C. Martin, E. André, G. Chollet, K. Karpouzis & D. Pelé (Eds.), *Intelligent Virtual Agents* (Vol. 4722, pp. 86-98): Springer Berlin / Heidelberg.
Noël, S., Dumoulin, S., & Lindgaard, G. (2009). Interpreting Human and Avatar Facial Expressions. In T. Gross, J. Gulliksen, P. Kotzé, L. Oestreicher, P. Palanque, R. Prates & M. Winckler (Eds.), *Human-Computer Interaction – INTERACT 2009* (Vol. 5726, pp. 98-110): Springer Berlin / Heidelberg.
Noel, S., Dumoulin, S., Whalen, T., & Stewart, J. (2006 ). Recognizing Emotions on Static and Animated Avatar Faces *IEEE International Workshop on Haptic Audio Visual Environments and their Applications, 2006. HAVE 2006.* (pp. 99 - 104 ).
Ochs, M., Niewiadomski, R., Pelachaud, C., & Sadek, D. (2005). Intelligent Expressions of Emotions. In J. Tao, T. Tan & R. Picard (Eds.), *Affective Computing and Intelligent Interaction* (Vol. 3784, pp. 707-714): Springer Berlin / Heidelberg.
Pelachaud, C., & Poggi, I. (2002). Subtleties of facial expressions in embodied agents. *The Journal of Visualization and Computer Animation, 13*(5), 301-312.
Picard, R. W. (1997). *Affective Computing*: MIT Press.
Picard, R. W., & Klein, J. (2002). Computers that recognise and respond to user emotion: theoretical and practical implications. *Interacting with Computers, 14*(2), 141-169.
Prendinger, H., & Ishizuka, M. (2005). THE EMPATHIC COMPANION: A CHARACTER-BASED INTERFACE THAT ADDRESSES USERS' AFFECTIVE STATES. *Applied Artificial Intelligence: An International Journal, 19*(3), 267 - 285.
Rosis, F. d., Pelachaud, C., Poggi, I., Carofiglio, V., & Carolis, B. D. (2003). From Greta's mind to her face: modelling the dynamics of affective states in a conversational embodied agent. *International Journal of Human-Computer Studies, 59*(1-2), 81-118.
Schmidt, K., Bhattacharya, S., & Denlinger, R. (2009). Comparison of Deliberate and Spontaneous Facial Movement in Smiles and Eyebrow Raises. *Journal of Nonverbal Behavior, 33*(1), 35-45.
Schmidt, K. L., Cohn, J. F., & Tian, Y. (2003). Signal characteristics of spontaneous facial expressions: automatic movement in solitary and social smiles. *Biological Psychology, 65*(1), 49-66.
Traum, D., Marsella, S., Gratch, J., Lee, J., & Hartholt, A. (2008). Multi-party, Multi-issue, Multi-strategy Negotiation for Multi-modal Virtual Agents *Intelligent Virtual Agents* (pp. 117-130).
Vinayagamoorthy, V., Gillies, M., Steed, A., Tanguy, E., Pan, X., Loscos, C., et al. (2006). Building Expression into Virtual Characters *EUROGRAPHICS 2006 State of the Art Reports*.
Vinayagamoorthy, V., Steed, A., & Slater, M. (2005). Building characters: Lessons drawn from virtual environments *Toward Social Mechanisms of Android Science at CogSci 2005* (pp. 119-126).
Yannakakis, G., & Hallam, J. (2006). Towards Capturing and Enhancing Entertainment in Computer Games *Advances in Artificial Intelligence* (Vol. 3955, pp. 432-442): Springer Berlin / Heidelberg.